\documentclass[conference,9pt]{IEEEtran}
\IEEEoverridecommandlockouts

\usepackage{cite}
\usepackage{amsmath,amssymb,amsfonts}
\usepackage{algorithmic}
\usepackage{graphicx}
\usepackage{textcomp}
\usepackage{xcolor}
\usepackage{multirow}
\usepackage{booktabs}
\usepackage{url}
\def\BibTeX{{\rm B\kern-.05em{\sc i\kern-.025em b}\kern-.08em
    T\kern-.1667em\lower.7ex\hbox{E}\kern-.125emX}}
\begin{document}

\title{Enhancing Expressive Voice Conversion with Discrete Pitch-Conditioned Flow Matching Model}

\author{\IEEEauthorblockN{Jialong Zuo\textsuperscript{\rm *}\thanks{\textsuperscript{\rm *} Equal contribution.}, Shengpeng Ji\textsuperscript{\rm *}, Minghui Fang\textsuperscript{\rm *}, Ziyue Jiang, Xize Cheng, Qian Yang,\\ Wenrui Liu, Guangyan Zhang, Zehai Tu, Yiwen Guo, Zhou Zhao$^{\dagger} $\thanks{$\dagger$ Corresponding author.}}
\IEEEauthorblockA{\textit{$^1$Zhejiang University, $^2$LightSpeed Studios, Tencent, $^3$Independent Researcher}}
\IEEEauthorblockA{$\left \{ jialongzuo, zhaozhou \right \}$ @zju.edu.cn}
}

\maketitle

\begin{abstract}
This paper introduces PFlow-VC, a conditional flow matching voice conversion model that leverages fine-grained discrete pitch tokens and target speaker prompt information for expressive voice conversion (VC). Previous VC works primarily focus on speaker conversion, with further exploration needed in enhancing expressiveness (such as prosody and emotion) for timbre conversion. Unlike previous methods, we adopt a simple and efficient approach to enhance the style expressiveness of voice conversion models. Specifically, we pretrain a self-supervised pitch VQVAE model to discretize speaker-irrelevant pitch information and leverage a masked pitch-conditioned flow matching model for Mel-spectrogram synthesis, which provides in-context pitch modeling capabilities for the speaker conversion model, effectively improving the voice style transfer capacity. Additionally, we improve timbre similarity by combining global timbre embeddings with time-varying timbre tokens. Experiments on unseen LibriTTS test-clean and emotional speech dataset ESD show the superiority of the PFlow-VC model in both timbre conversion and style transfer. Audio samples are available on the demo page \url{https://speechai-demo.github.io/PFlow-VC/.}

\end{abstract}


\section{Introduction}
Voice conversion (VC) is a technique for transforming speech signals, altering the original speaker's voice to sound like a target speaker while preserving the linguistic content. The general trend follows the analysis-resynthesis paradigm, where in the analysis phase, various disentanglement techniques are used to decompose the speech signal into different components, including speaker-independent content representation and speaker-related variations. During the resynthesis phase, the disentangled linguistic content, conditioned by the target speaker's representation, is integrated into the synthesis network to generate the converted audio.

Much research focuses on learning better linguistic representations or speaker-disentangled representations during the analysis stage. Information bottleneck methods are introduced into autoencoder structure to disentangle speaker and content~\cite{qian2019autovc,lee2021voicemixer}. AdaIN-VC \cite{chou2019one} uses instance normalization and AGAIN-VC \cite{chen2021again} additionally uses an activation function to constrain the speaker variations flowing from source speech. Some works~\cite{qian2020unsupervised,choi2021neural,ning2023expressive} process the original audio signal and adopt information perturbation methods to pertub the speaker characteristics. However, while these methods aim for effective speaker disentanglement, which may also harm the linguistic information in some degree \cite{qian2022contentvec}, resulting in low quality speech. Recently, advanced methods~\cite{kim2023unitspeech,van2022comparison,cheng2023opensr,choi2023diff,cheng2023transface} have directly utilized SSL features extracted from pre-trained self-supervised speech representation networks~\cite{hsu2021hubert,babu2021xls} as linguistic content. These features offer rich semantic information and exhibit minimal speaker variance \cite{van2022comparison}, leading to promising results in voice conversion tasks.

\begin{figure*}[t]
\centering
\includegraphics[height=6cm, width=17cm]{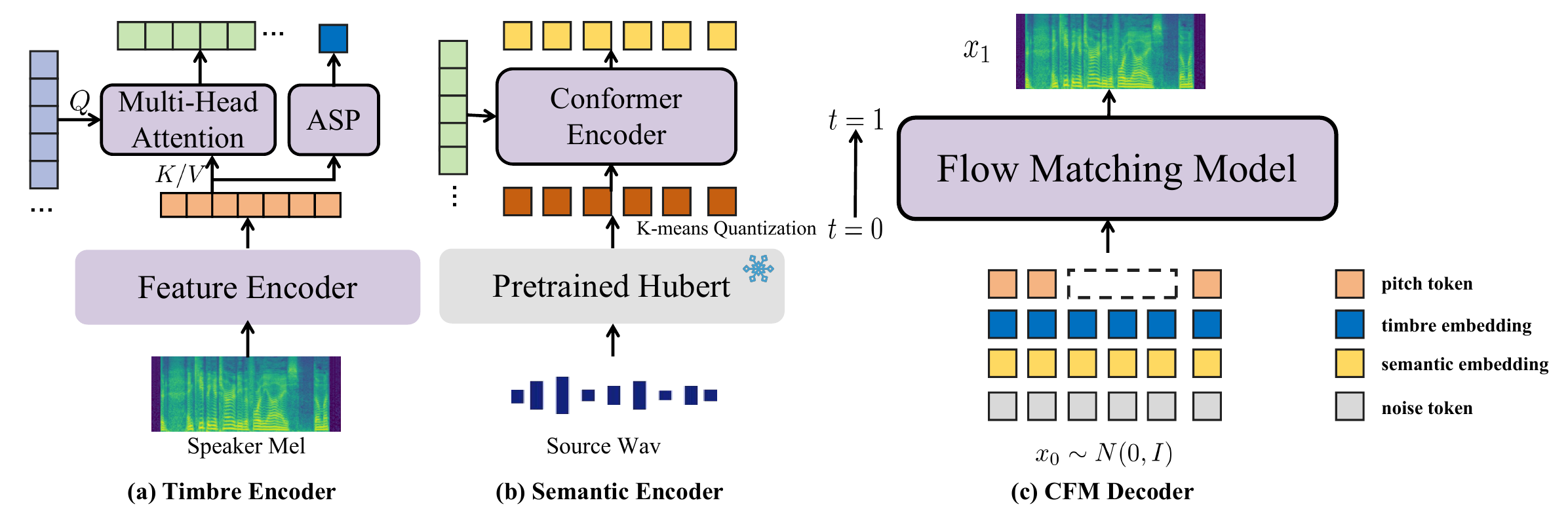}
\caption{The overall architecture of PFlow-VC.}
\label{overview}
\end{figure*}

 Another research avenue delves into the enhancement of speaker identity modeling during the synthesis stage. Methods based on speaker embeddings \cite{xiao2022dgc, cooper2020zero} typically leverage global speaker representations obtained from encoder modules or pre-trained speaker verification networks. However, these methods often falter in zero-shot voice conversion scenarios due to the inherent generalization limitations of global embeddings. Recent studies have explored more efficient speaker modeling techniques. For instance, SEF-VC \cite{li2024sef} employs a position-agnostic cross-attention mechanism to assimilate and incorporate speaker timbre from reference speech. Similarly, RefXVC \cite{zhang2024refxvc} utilizes both global and local speaker embeddings to capture timbre variations during speech conversion. Despite these advancements, many voice conversion (VC) methodologies remain predominantly focused on preserving the speaker's identity, leaving substantial room for further investigation into enhancing expressiveness, such as prosody and emotional nuances, during speaker conversion. RefXVC, for instance, replicates the prosody of the source speaker during synthesis, thereby lacking the capacity for style transfer. Conversely, other existing works~\cite{wang2022zero, yao2024promptvc, jiang2023fluentspeech} employ prosody predictors and encoders to derive prosody representations from Mel spectrograms, aiming to generate more natural-sounding speech. However, these implicit learning approaches suffer from a lack of interpretability, with the predicted prosody often being heavily influenced by the source content representation rather than the target speaker, resulting in diminished prosody transferability. Diff-HierVC \cite{choi2023diff} introduces a pitch generator based on a diffusion process to estimate F0 values tailored to the target voice style, achieving superior transfer performance. Nonetheless, the complexity of multiple diffusion models significantly impedes inference speed.

To address these challenges, we propose PFlow-VC, a simple and efficient discrete pitch-conditioned flow matching model for expressive voice conversion tasks. We introduce the use of a pretrained pitch VQVAE model to discretize pitch information, extracting speaker-independent prosody representations. Additionally, we employ a multi-scale timbre encoder to model global and time-varying timbre variations, and a pretrained self-supervised HuBERT model for semantic token extraction. We then utilize a powerful mask conditional flow matching model to reconstruct Mel spectrograms from these disentangled representations. On the one hand, the mask pitch-conditioned strategy provides in-context pitch modeling capabilities for the speaker conversion model, thereby ensuring that the synthesized speech is prosody-consistent with the target speaker. On the other hand, we propose to use time-invariant speaker characteristics and time-varying timbre tokens for better timbre similarity. Moreover, thanks to the rectified flow matching technique, the model can generate high-quality audio with fewer sampling steps. Comprehensive evaluations reveal that PFlow-VC achieves significantly improved capacity for zero-shot voice conversion and expressive voice conversion. Ablation studies further demonstrate the effectiveness of pitch discrete token conditioning and timbre modules.


\section{PFlow-VC Architecture}
\subsection{Architecture Overview}
 The overall architecture is illustrated in Figure \ref{overview}. For expressive voice conversion, we define speech in three components: linguistic content, timbre variations, and quantized pitch representation. First, we pretrain a pitch VQVAE model using speaker-mean normalized log F0, representing prosodic features as discrete random variables. For timbre modeling, we introduce a timbre encoder that extracts both timbre-invariant and timbre-varying speaker embeddings from the source speech. We utilize a pretrained HuBERT model and a k-means model to obtain discrete semantic tokens, which serve as content features and exhibit minimal speaker variance. The extracted semantic tokens are fed into a Conformer with a cross-attention layer and time-varying timbre embedding to obtain the semantic embedding, as shown in Figure \ref{overview} (b). A generative conditional flow matching model is then used to synthesize Mel spectrograms from these disentangled features. Finally, a pretrained HiFi-GAN vocoder \cite{kong2020hifi} synthesizes a perceptible waveform from the generated Mel spectrogram.
\subsection{Pitch Quantizer}
Generally, prosodic information includes both duration and pitch. In our PFlow-VC model, we directly ensure that the duration of the generated speech matches the source speech, so we primarily focus on pitch representation. Many works use fundamental frequency (F0) to encode prosodic information; however, it also encodes a significant amount of speaker information. Therefore, we propose to model \textbf{S}peaker-\textbf{M}ean \textbf{N}ormalized log F0 with a VQVAE model.
\begin{equation}
\text{SMN-logf0} = \log f - \mathbb{E}_{f’ \in speaker}[\log f’]
\end{equation}
which can be interpreted as the ratio to the mean pitch in the log space. Specifically, the equation above is used for voiced frames, and the expectation is taken over voiced frames from a speaker. For unvoiced frames, we simply set $f0 = 0$. We then adopt the VQ-VAE framework to generate discrete pitch representations, following \cite{polyak2021speech,ji2024wavtokenizer,fang2024ace}. The VQ-VAE employs a convolutional encoder, a bottleneck with a learned codebook $C = (e_{1}, \ldots, e_{m})$, where each item in $C$ is a 128-dimensional vector, and a decoder for original SMN-f0 reconstruction. The quantized indices extracted by the VQ layer serve as the final discrete pitch token representation. To ensure more stable training, we use Exponential Moving Average (EMA) updates to learn the codebook and employ random restarts for unused embeddings.
\subsection{Timbre Encoder}
Prior research often use an averaged speaker embedding, which fails to capture the dynamic nature of timbral characteristics. Additionally timber is not static but rather various over time, depending on time-varying contents such as linguistic information and pitch. Therefore, we define speaker representations as two components: global and time-varying timbre tokens, similar to \cite{zhang2024refxvc,jiang2023mega}. As shown in Figure \ref{overview} (a), our timbre encoder is based on the architecture in \cite{desplanques2020ecapa}. The feature encoder consists of ECAPA-TDNN and Multilayer Feature Aggregation blocks. Timbre tokens are extracted via a cross-attention mechanism, with 64 trainable latent vectors as queries and the encoder's output as keys and values. We also obtain a global speaker embedding through attentive statistical pooling (ASP). A conformer encoder is used to encode the discrete semantic tokens, with a cross-attention layer between the self-attention and convolution modules to integrate time-varying timbre embeddings. This approach aims to accurately characterize dynamic speaker information, enhancing the preservation of speaker identity.

\subsection{Mask Conditional Flow Matching Model}
Continuous Normalizing Flows (CNFs) \cite{chen2018neural} aims to estimate the unknown distribution $q(x)$ of data $x \in \mathbb{R}^d$ by learning the probability path from a simple prior distribution $p_0$ to a data distribution $p_1 \approx q$. This mapping can be further taken as a time-dependent changing process of probability density (a.k.a. flow), determined by the ODE:
\begin{equation}
\frac{d}{dt}{\phi_t(x)} = v_t(\phi_t(x)); \quad \phi_0(x) = x.
\end{equation}
where $v_t : [0, 1] \times \mathbb{R}^d \rightarrow \mathbb{R}^d$ is a vector field that generates the flow $\phi_t : [0, 1] \times \mathbb{R}^d \rightarrow \mathbb{R}^d$. We can sample from the approximated data distribution $p_1$ by solving the initial value problem in Eq. (1). Suppose there exists a known vector field $u_t$ that generates a probability path $p_t$ from $p_0$ to $p_1$. The flow matching loss is defined as:
\begin{equation}
L_{FM}(\theta) = E_{t,p_t(x)}\|u_t(x) - v_t(x; \theta)\|^2.
\end{equation}
where $t \sim U[0, 1]$ and $v_t(x; \theta)$ is a neural network with parameters $\theta$. However, $L_{FM}$ is uncomputable for lack of prior knowledge of $p_t$ or $v_t$. Luckily, \cite{lipman2022flow} proposed Conditional Flow Matching (CFM) objective presented as: 
\begin{equation}
L_{CFM}(\theta) = E_{t,q(x_1),p_t(x|x_1)}\|u_t(x|x_1) - v_t(x; \theta)\|^2
\end{equation}
By conditioning $p_t$ and $v_t$ on real data $x_1$, FM and CFM have identical gradients with respect to $\theta$ for training generative model.

In PFlow-VC, an optimal-transport conditional flow matching model (OT-CFM) \cite{lipman2022flow} is employed to learn the distribution of Mel spectrogram and generate samples from it with generated tokens as conditions. The OT-CFM loss function can be written as:
\begin{equation}
L(\theta) = E_{t,q(x_1),p_0(x_0)}\|u^{OT}_t(\phi^{OT}_t(x)|x_1) - v_t(\phi^{OT}_t(x)|\theta)\|^2
\end{equation}
by defining $\phi^{OT}_t(x) = (1 - (1 - \sigma_{\text{min}})t)x_0 + tx_1$ as the flow from $x_0$ to $x_1$ where each datum $x_1$ is matched to a random sample $x_0 \sim N(0, I)$. The global speaker embedding $e_s$, semantic tokens $\{\mu_t\}_{1:T}$, and masked pitch token $\tilde{p}_1$ are also fed into the our decoder network to match the vector field with learnable parameters $\theta$: 
\begin{equation}
v_t(\phi^{OT}_t(x)|\theta) = f_{\theta} \left( \phi^{OT}_t(x), t; e_s, \{\mu_t\}_{1:T}, \tilde{p}_1 \right)
\end{equation}
During training, we randomly mask a certain proportion of pitch tokens to force the model to learn the masked pitch information based on the context of the unmasked pitch representations. This enables the model to develop in-context pitch modeling capabilities, which effectively transfer the prosody of the target speaker during inference.

\subsection{Trainging and Inference}
During training, we randomly mask continuous segments of pitch tokens to implement contextual pitch prompting. At inference, we first concatenate the target speaker's semantic tokens, pitch tokens, and speaker embedding along the channel dimension, then prepend them to the tokens to be synthesized. The source speaker's corresponding pitch tokens are replaced with null tokens and fed into the conditional flow matching decoder. Additionally, we adopt Classifier-free Guidance (CFG) for our generative model following \cite{du2024cosyvoice}. 

\section{Experiments}
\subsection{Experimental Setup}
\textbf{Datasets} We train our PFlow-VC and various baselines on LibriTTS \cite{zen2019libritts}, a large-scale multi-speaker English corpus of approximately 585 hours. We evaluate timbre similarity on the test-clean subset (from LibriSpeech \cite{panayotov2015librispeech}) and style transfer capability on the ESD dataset \cite{esd}. For speaker conversion experiment, we randomly select 20 speakers: 10 for source speech and 10 for target speech, forming a 200-sample test set. For style conversion experiment, we use the utterances of 10 speakers as target speech, creating another 200-sample test set. Notably, the timbres in both test-clean and ESD datasets are unseen and do not overlap with the training data.\\
\textbf{Training Setup}
Each recording is sampled at 16000 Hz with 16-bit quantization. The ground truth mel-spectrograms are generated from the raw waveform with a frame size of 1024, a hop size of 256, and 80 channels. We trained a HiFi-GAN vocoder using the same LibriTTS training data. We extracted 1024-dimensional semantic features from a pretrained HuBERT model \footnote{\url{https://github.com/facebookresearch/textlesslib}} trained on 960 hours of LibriSpeech. These semantic features were then quantized offline using K-Means clustering with 500 centers. For training the Pitch VQVAE, we use a single VQ codebook with a size ($C$) of 64 and a vector dimension of 128. The coefficients for pitch reconstruction, commitment loss, and VQ loss are set to 1.0, 0.15, and 0.05, respectively. The structure of our flow matching decoder is similar to \cite{mehta2024matcha}.\\
\textbf{Evaluation Details}
In our experimental analysis, we use both objective and subjective metrics to evaluate synthesis quality, speaker similarity, and emotion consistency. Objective metrics include word error rate (WER), speaker similarity (SECS), and emotion accuracy between output speech and prompts. Subjective evaluation involves MOS (mean opinion score) assessments for audio naturalness, including QMOS (quality, clarity, and naturalness) and SMOS (speaker similarity).\\
\textbf{Baseline Models}
We compared PFlow-VC with the following advanced VC methods:
\begin{itemize}
    \item \textbf{YourTTS}, a speaker-embedding based end-to-end VC model that performs speaker disentanglement using normalizing flows.
    \item \textbf{Diff-HierVC}, a hierarchical VC system based on two diffusion models with robust pitch generation.
    \item  \textbf{SEF-VC}, a speaker embedding-free voice conversion model utilizing a position-agnostic cross-attention mechanism.
\end{itemize}
All the baseline models in our experiment are trained on the same training set for fair comparison.

\subsection{ Zero-shot VC Results}
We evaluated the model's performance in a zero-shot voice conversion scenario on an unseen test-clean dataset consisting of 200 samples. For the WER metric, we used the Whisper-Large model \footnote{\url{https://huggingface.co/openai/whisper-large-v3}} to transcribe the synthesized speech. Additionally, we calculated speaker cosine similarity (SECS) using a pretrained speaker verification model \footnote{\url{https://huggingface.co/microsoft/wavlm-base-plus-sv}}. The experimental results are shown in Table \ref{table1}. 

Both YourTTS and Diff-HierVC use global representation for speaker identity modeling, which lacks sufficient generalization, making it difficult to handle zero-shot speaker conversion. Thanks to a position-agnostic cross-attention mechanism, SEF-VC achieves effective timbre conversion, significantly improving speaker similarity. However, the perceived quality of the synthesized audio is somewhat lacking, likely due to limitations in the decoder's performance. In unseen scenarios, the low correlation between source content and target speaker mel content can substantially impair the learning capability of the cross-attention mechanism. This results in the acquisition of redundant information, which negatively impacts the quality of the generated audio. Compared to these systems, PFlow-VC excels in modeling and learning timbre by combining global embeddings with time-varying timbre tokens. Trainable latent vectors capture dynamic speaker characteristics, while the powerful generative ability of the conditional flow matching model produces high-quality speech. This approach achieves superior or competitive results in speaker similarity and audio quality metrics for zero-shot voice conversion tasks. 

\begin{table}[ht]
\centering
\tabcolsep=3.5pt
\caption{Comparison of different systems for zero-shot VC.}
\begin{tabular}{l|ccccc}
\toprule
\bfseries Method & \bfseries {WER$\downarrow$} & \bfseries {SECS$\uparrow$} & \bfseries QMOS $\uparrow$ & \bfseries SMOS $\uparrow$ & \bfseries RTF $\downarrow$ \\ 
\midrule
Source GT &2.090&-&4.28 $\pm$ 0.06&-&-\\
\midrule
Your-TTS  &5.818&0.867&3.68 $\pm$ 0.08&3.75 $\pm$ 0.07&\bfseries0.07\\
Diff-HierVC  &2.607&0.905&\bfseries{4.07 $\pm$ 0.11}&4.03 $\pm$ 0.09&0.42\\
SEF-VC  &3.024&0.911&3.92 $\pm$ 0.09&4.12 $\pm$ 0.10&0.08\\
PFlow-VC &\bfseries2.574&\bfseries0.920&4.05 $\pm$ 0.08&\bfseries4.20 $\pm$ 0.07&0.13\\
\bottomrule
\end{tabular}
\label{table1}
\end{table}

\begin{table}[t]
\centering
\caption{The emotion style consistency between the synthesized audio and the target speaker speech.}
\begin{tabular}{l|ccc}
\toprule
\bfseries Method & \bfseries {WER$\downarrow$} & \bfseries {SECS$\uparrow$} & \bfseries Emo-Consistency Score $\uparrow$ \\ 
\midrule
Target Spk &2.130&-&0.937\\
\midrule
Your-TTS  &6.740&0.849&0.512\\
Diff-HierVC  &3.033&0.873&0.650\\
SEF-VC  &3.550&0.886&0.622\\
PFlow-VC &\bfseries2.928&\bfseries0.903&\bfseries0.725\\
\bottomrule
\end{tabular}
\label{table2}
\end{table}
Furthermore, by leveraging the reflow characteristics of conditional flow matching, PFlow-VC excels in producing high-quality speech with expedited synthesis times. As evidenced in Table \ref{table1}, although the model does not outperform end-to-end VC systems (YourTTS and SEF-VC) in terms of the Real-Time Factor (RTF) metric, it still delivers commendable results which significantly enhances inference speed compared to Diff-HierVC.
\subsection{ Emotion Style Transfer Ability Evaluation}
\begin{figure}[th]
\centering
\includegraphics[height=4cm, width=8cm]{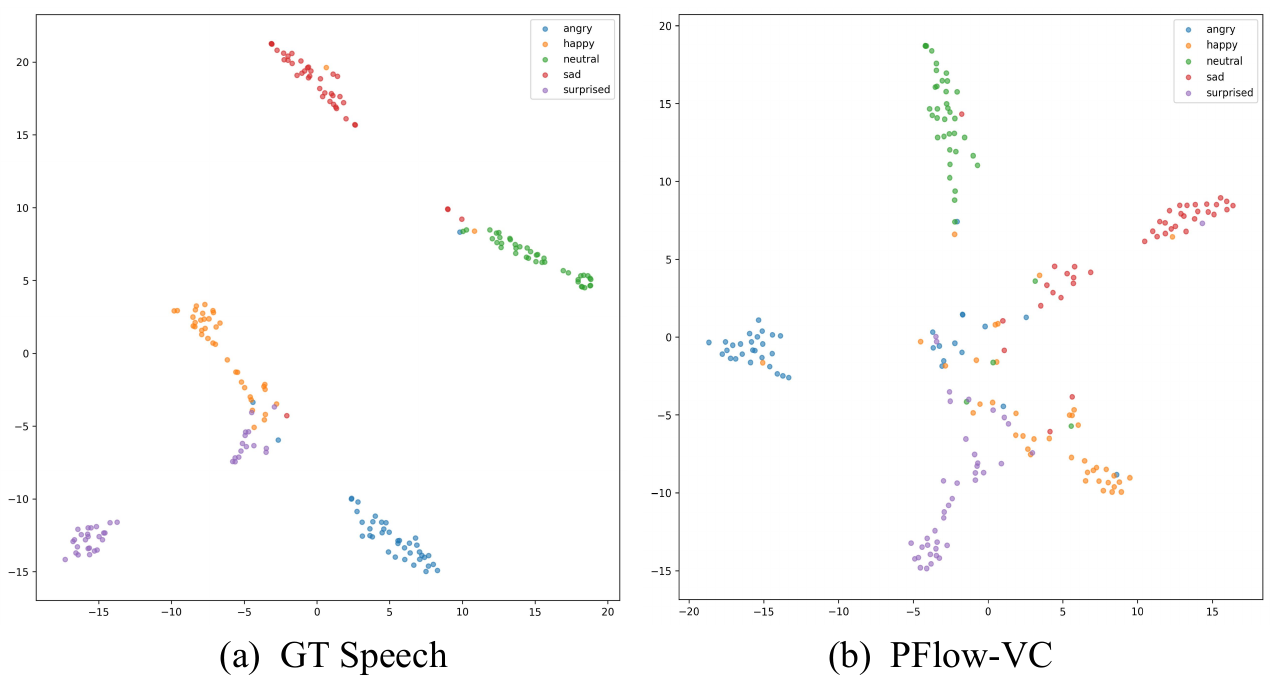}
\caption{The visualization of emotion representations extracted from unseen ESD test set.}
\label{emo_embed}
\end{figure}

To validate the capacity of PFlow-VC to enhance voice conversion expressiveness or voice style transfer, we combined source speakers from the test-clean dataset with the out-of-domain emotional dataset ESD to create an emotional test set. Utilizing a pretrained emotion2vec \cite{ma2023emotion2vec} model as an emotion classifier \footnote{\url{https://github.com/ddlBoJack/emotion2vec}}, we calculated the emotion score for each synthesized speech sample. These scores were then averaged to obtain the Emo-Consistency Score. As shown in Table \ref{table2}, PFlow-VC outperforms the baseline in WER and SECS metrics and significantly exceeds other models in emotion classification scores. This demonstrates our model's ability to learn and transfer emotional states from target speech to source speech, maintaining consistent emotions. We attribute this to the introduction of the self-supervised discrete pitch token conditioning and the mask-based contextual training strategy, which enable the model to learn pitch distribution from prompt speech. Combined with the denoising process of the conditional flow matching decoder, this results in more emotionally consistent speech. Additionally, as depicted in Figure \ref{emo_embed}, we used t-SNE to visualize the clustering of emotion representations from the emotion2vec model on the emotional test set generated by PFlow-VC, confirming that the synthesized speech effectively conveys the intended emotions.

\subsection{Ablation Studies}
\subsubsection{Pitch Modeling}
We conducted an ablation study to investigate the impact of discrete pitch representation on the performance of the voice conversion model. In our PFlow-VC architecture, we pre-trained a pitch VQVAE model to quantize pitch tokens at 25Hz per second. These pitch tokens were then used as an additional input to the conditional flow matching model, employing a mask training strategy to implicitly model pitch distribution within an utterance. During inference, the model utilizes pitch prompt information from the target speaker to guide the pitch variations of the source utterance.

\begin{table}[th]
\centering
\caption{Ablation Studies on pitch modeling and speaker representation modeling. $\textbf{tv}$ means time-varying, and \textbf{CFM-VC} indicates that only Hubert semantic tokens and global speaker embeddings are used.}
\begin{tabular}{l|cccc}
\toprule
\bfseries Method & \bfseries {WER$\downarrow$} & \bfseries {SECS$\uparrow$} & CMOS-Q & CMOS-S \\ 
\midrule
PFlow-VC &2.574&0.920&0.00&0.00\\
\midrule
w/o $pitch\,token$ &  3.014&0.907&-0.11&-0.13\\
w/o $tv\,timbre\,token$ &  2.706&0.877&-0.03&-0.41\\
w/i $corss\,attention$ &  2.810&0.908&-0.05&-0.11\\
CFM-VC &3.230&0.865&-0.16&-0.47\\
\bottomrule
\end{tabular}
\label{table3}
\end{table}

Compared to previous methods that directly perform F0 transformation with denormalization \cite{qian2020f0} or predict F0 using an F0 Encoder, our approach enables the model to contextually model pitch representation, resulting in more natural synthesis. Previous work has demonstrated that models supervised with pitch loss for prediction often produce generated prosody that sounds dull and monotonous \cite{kharitonov2021text}. Our method addresses this issue by allowing the model to adaptively capture and reproduce the dynamic pitch variations present in natural speech, thus achieving more expressive and lifelike voice conversion. It is clear from Table \ref{table3} that the absence of pitch token conditioning leads to a noticeable decline in both audio quality and timbre similarity metrics. This highlights the essential role of pitch tokens in ensuring high-quality voice conversion.
\subsubsection{Speaker Representations}
As is well known, preserving speaker identity is crucial in voice conversion models. In our PFlow-VC model, we utilize both global speaker embeddings and time-varying timbre tokens to enhance timbre modeling capabilities and improve generalization in zero-shot scenarios. To validate the effectiveness of this timbre encoder, we compared our approach with methods using only speaker embeddings and cross-attention speaker information prompting. As shown in Table \ref{table3}, when relying solely on speaker embeddings to represent speaker identity, the timbre cloning performance on the test set significantly decreases. A key reason is that a single embedding cannot characterize dynamic speaker information that varies over time in speech signals. When replacing the global speaker encoder with a cross-attention mechanism similar to \cite{li2024sef}, the timbre similarity metrics exhibit substantial improvement. This method enables the model to better capture the evolving nature of speaker characteristics, leading to more accurate timbre preservation. However, it is important to note that even with this improvement, the performance still falls short of the effectiveness achieved by the timbre encoder in PFlow-VC. Our timbre encoder utilizes dynamic timbre tokens, which offer a more detailed and adaptive representation of speaker traits over time. We believe that this design significantly boosts the model's robustness in timbre modeling.

\section{Conclusions} 

We introduce PFlow-VC, an advanced conditional flow matching model designed to enhance expressiveness in voice conversion. By integrating fine-grained discrete pitch tokens and target speaker prompts, PFlow-VC effectively addresses the limitations of current
methods in capturing prosodic details. The model employs a self-supervised pitch VQVAE to discretize speaker-independent pitch information and utilizes a masked pitch-conditioned flow matching mechanism for Mel-spectrogram synthesis. This significantly improves timbre conversion and style transfer. Experiments on unseen datasets validate the model’s superior performance in achieving expressive and natural voice conversion.

\newpage

\bibliographystyle{IEEEtran}
\bibliography{refs}
\end{document}